\documentclass[a4paper]{article}
\usepackage{amscd,amsmath,amssymb,enumerate}
\usepackage{xcolor}

\newcommand{\p}[1]{(\ref{#1})}

\newcommand{\cS}{{\cal S}}

\newcommand{\hz}{{\hat z}}

\newcommand{\be}{\begin{equation}}
\newcommand{\ee}{\end{equation}}
\newcommand{\bea}{\begin{eqnarray}}
\newcommand{\eea}{\end{eqnarray}}

\newcommand{\ba}{\begin{array}} \newcommand{\ea}{\end{array}}

\def\im{{\rm i\,}}

\newcommand{\nn}{\nonumber}

\begin{document}
\begin{flushright}
\end{flushright}\vspace{1cm}
\begin{center}
{\Large\bf On the origin of  Higher Schwarzians}
\end{center}
\vspace{1cm}

\begin{center}
{\Large\bf 
S.~Krivonos}
\end{center}

\vspace{0.2cm}

\begin{center}

{\it
 Bogoliubov  Laboratory of Theoretical Physics, JINR,
141980 Dubna, Russia \\
and \\
Laboratory of Applied Mathematics and Theoretical Physics, TUSUR, Lenin. Ave 40, 634050 Tomsk, Russia}

\vspace{0.5cm}

{\tt  krivonos@theor.jinr.ru}
\end{center}
\vspace{2cm}

\begin{abstract}
\noindent In this paper we analyze  higher Schwarzians and show that they 
are closely related to the nonlinear realization of the Virasoro algebra. The Goldstone fields of
such a realization provide a new set of  $SL(2,\mathbb{R})$ invariant higher Schwarzians that are deeply related to the Aharonov \cite{aharonov}, Tamanoi \cite{tamanoi} and Bonora-Matone \cite{BM} ones. A minor change of the coset space parametrization leads to a new set of $SL(2,\mathbb{R})$ non-invariant higher Schwarzians now related to the Schippers \cite{schippers} and Bertilsson \cite{bert} Schwarzians.
\end{abstract}

\vskip 1cm
\noindent
PACS numbers: 11.30.Pb, 11.30.-j

\vskip 0.5cm

\noindent
Keywords: Schwarzian, conformal symmetry, Virasoro algebra

\newpage

\setcounter{equation}{0}
\section{Introduction. Standard Schwarzian within nonlinear realization}
The Schwarzian derivatives or simply Schwarzians has been known for a long time. Their story goes back to Lagrange who introduced a version of the Schwarzians (see e.g. \cite{OT}  for a
survey). Later on, this derivative appeared in  projective and conformal geometry as well as in many other contexts in mathematics and mathematical physics. 

The Schwarzian derivative $S_t$ is defined by the relation
\be\label{SchwDer}
S_t = \frac{\dddot t}{\dot t} - \frac{3}{2} \left( \frac{\ddot t}{\dot t}\right)^2  =\frac{d}{d \tau }\left( \frac{\ddot t}{\dot t}\right) -
\frac{1}{2} \left( \frac{\ddot t}{\dot t}\right)^2 = \dot{T}_t-\frac{1}{2} T_t^2
, \;\; \dot t = \partial_\tau t.
\ee
Here, $T_t={\ddot t}/{\dot t}$ is the pre-Schwarzian derivative of $t(\tau)$.
Applications of the Schwarzian derivative are especially connected with problems of univalent analytic functions \cite{ aharonov, tamanoi, schippers, bert,nehari, kim}.
Despite its well-established role in mathematical physics, the Schwarzian derivative
remains somewhat mysterious. It is not a derivative, exactly, but what is it? 
The most known property of the Schwarzian derivative is its invariance under $SL(2,\mathbb{R})$ transformations, acting on $t$, i.e.
\be\label{eq1}
\mbox{if   } t^\prime = \frac{a t +b}{c t + d},\quad a d- b c\neq 0 \;\;\quad \mbox{then}\quad  \;\; S_{t^\prime} = S_t.
\ee
The infinitesimal form of the condition \p{eq1} is
\be\label{eq2}
\delta t=a_{-1}+a_0 t+\frac{a_1}{2} t^2 , \qquad S_t[t+\delta t ]=S_t[t].
\ee
The parameters $a_{-1},a_0,a_1$ correspond to translation, dilatation, and conformal boost, respectively. The group $SL(2,\mathbb{R})$ with the commutative relations
\be\label{sl2}
\im \left[ L_n, L_m\right] = (n-m) L_{n+m}, \qquad n,m= -1,0,1
\ee 
is just one dimensional  conformal group.

It is interesting, but pre-Schwarzian derivative is not invariant under $SL(2,\mathbb{R})$ transformations
\be\label{eq3}
T_t[t+\delta t] =  T_t[t] + a_1  \dot{t} .
\ee
As we will see below \p{eom1}, the $\dot{t}$ starts from a constant, and therefore
the  pre-Schwarzian derivative $T_t$ is shifted by a constant under the transformation
$\delta t = a_1 t^2$. Thus, $T_t$ is the Goldstone field accompanying the spontaneous
breaking of the conformal boost. On the other hand, we have
\be\label{eq4} 
\delta Log[\dot{t}] = a_0 +  a_1 t ,
\ee
and thus $Log[\dot{t}] $ is the Goldstone field for the spontaneously broken dilatation.
Finally, note that the translations $\delta t= a_{-1}$ are also spontaneously broken with
$t(\tau)$ being the corresponding Goldstone field.

Within a nonlinear realization of  spontaneously broken $sl(2,\mathbb{R})$ symmetry,
the natural place for the Schwarzian is to be associated with one of the $sl(2,\mathbb{R})$ Cartan forms $\omega_{-1},\omega_{0},\omega_1$ defined in the standard way as
\be\label{CF}
g^{-1} dg = \im \omega_{-1} L_{-1} +  \im \omega_{0} L_{0} + \im \omega_{1} L_{1},
\qquad g=e^{\im t L_{-1}} e^{\im u L_0}  e^{ \frac{\im}{2} z_1 L_1} .
\ee
Explicitly, these forms read
\be\label{CF1}
\omega_{-1} = e^{-u} dt, \quad \omega_0 = du - e^{-u} z_1 dt, \quad \omega_1 = \frac{1}{2}\left[ dz_1 - z_1 du + \frac{1}{2} e^{-u} z_1^2 dt\right]  .
\ee
All these forms are invariant with respect to  $sl(2,\mathbb{R})$ symmetry \p{eq2}. Thus, one can
introduce invariant time $\tau$ and reduce the number of independent fields (Inverse Higgs phenomenon \cite{IO}) by imposing the following constraints:
\be\label{con1}
e^{-u} dt = d\tau, \qquad \omega_0 = du -  e^{-u} z_1 dt =0.
\ee
As a result of these constraints \p{con1} and treating all fields as dependent on the invariant time, $\tau$ one can get
\be\label{eom1}
\dot{t} = e^{u(\tau)}, \quad (a) \qquad z_1 = \dot{u} = \frac{\ddot{t}}{{\dot t}} = T_t \quad (b).
\ee
Substituting expressions in \p{eom1} into form $\omega_1$, we  have
\be\label{eom2}
\omega_1 =\frac{1}{2} d\tau\left[  \frac{d}{d \tau }\left( \frac{\ddot t}{\dot t}\right) -
\frac{1}{2} \left( \frac{\ddot t}{\dot t}\right)^2\right] = \frac{1}{2} S_t d\tau .
\ee 
Thus, as was expected, the Schwarzian $S_t$  appears as the $d\tau$-projection of the Cartan form $\omega_1$ associated with the generator of the conformal boost $L_1$. 

This approach to  Schwarzians was initiated by A.~Galajinsky in \cite{AG1}. Later, on it was generalized to the cases of supersymmetric Schwarzians in \cite{AG2,AG3,GK,KK1,KK2,KK3}.
Being very productive, this approach can tell nothing about higher Schwarzians: $sl(2,\mathbb{R})$ invariant higher Schwarzians of D.~Aharonov \cite{aharonov} as well as  higher Schwarzians of H.~Tamanoi \cite{tamanoi} were completely out of game. The nature of $sl(2,\mathbb{R})$ non-invariant higher Schwarzians of E.~Schippers \cite{schippers} and D.~Bertilsson \cite{bert} was also unclear. The goal of the present paper is to provide a unified 
description of $sl(2,\mathbb{R})$ invariant higher Schwarzians as Goldstone fields for spontaneously broken Virasoro symmetry spanned by the generators $\left\{ L_n, n\geq -1\right\}$. From this point of view, non-invariant higher Schwarzians are just specific deformations of invariant ones by $T_t$ dependent terms.

\setcounter{equation}0
\section{Higher Schwarzians }

In 1969, D.~Aharonov gave definitions of higher-order analogues of the Schwarzian derivative
\cite{aharonov}. Later on, Tamanoi introduced another set of higher order Schwarzian derivatives
\cite{tamanoi}. Finally, Kim and Sugawa derived relations between the Aharonov invariants and Tamanoi's Scwarzian derivatives \cite{kim}. Another, more physical definition of the higher Schwarzians\footnote{The Author is grateful to M.Matone for drawing his attention to the papers \cite{BM}.}  has been formulated by L.~Bonora and M.~Matone \cite{BM}.

All these definitions lead to higher order Schwarzian derivatives invariant with respect to $sl(2,R)$ transformations \p{eq1}, \p{eq2}  and both definitions are non-geometric ones. Using the results from the previous Section, one can propose a purely geometric definition of higher Schwarzians. The basic idea comes from the transformation of the Schwarzian under transformation
\be\label{L2}
\delta t =\frac{a_2}{6} t^3 \quad \rightarrow \quad \delta S_t = a_2\, t'[\tau]^2 = a_2 + \ldots.
\ee
Thus, we see that the Schwarzian itself behaves like a Goldstone boson for partially broken symmetry \p{L2}. It is not too difficult to verify that transformation \p{L2} together with \p{eq2} form the centerless subalgebra of the Virasoro algebra
\be\label{Vir}
\im \left[ L_n, L_m \right] = (n-m) L_{n+n}, \qquad n,m \geq -1 .
\ee
It is a natural guess to associate the Schwarzian with the Goldstone fields for transformation \p{L2} generated by the operator $L_2$. With such identification, one can expect that higher Schwarzians will appear as the Goldstone fields associated with the higher generators $L_n , n \geq 3$. Technically, the realization of this idea consists in three steps: 
\begin{itemize}
\item{Step 1}\\
One has to choose a proper parametrization of the group element corresponding to the algebra \p{Vir} 
\be\label{g}
g= e^{\im\, t L_{-1}} e^{\im\, u L_0}\;  e^{\frac{\im}{2}\, z_1 L_1} \prod_{i=2}^{\infty} e^{\frac{\im}{(i+1)!} z_i L_i}
\ee
The order of the first three exponents is fixed by the invariance of the Goldstone field\footnote{We remember that the Goldstone $z_1$ has to be identified with the
pre-Schwarzian $T_t$.} $z_1$ under the transformation generated by $L_0$ (dilatations) \p{eq3}.
\item{Step 2}\\
Next, one has to calculate the Cartan forms
\be\label{cf1}
\Omega = g^{-1} d g = \im\; \sum_{n=-1} \omega_n L_n .
\ee
 Let us list several first Cartan forms :
\bea\label{cf2}
\omega_{-1} & = & e^{-u} dt, \nn \\
\omega_{0} & = & du -  z_1 e^{-u} dt , \nn \\
\omega_1 & = &\frac{1}{2} \left[  dz_1 -z_1 \, du -e^{-u}\left(z_2  -\frac{1}{2} z_1^2 \right)\,dt \right] , \nn \\
\omega_2 & = &\frac{1}{6}\left[ d z_2 - 2 z_2\, du - e^{-u}\left( z_3 -2\,  z_1\, z_2 \right)\, dt, \right] ,\nn \\
\omega_3 & = & \frac{1}{24}\left[ dz_3 -2\, z_2\, dz_1 +2\, du\, z_1\, z_2 - 3\, du\, z_3 -e^{-u} \left( z_4 - 3\, z_1\, z_3 -z_2^2 + z_1^2\, z_2\right)\, dt \right] , \nn \\
\omega_4 & = & \frac{1}{120}\left[ dz_4 - 5  z_1  z_3 + 5 du z_1 z_3 - 4 du z_4 - e^{-u} \left( z_5- 4 z_1 z_4 -5 z_2 z_3 + \frac{5}{2} z_1^2 z_3\right)\,dt \right],\quad etc.
\eea
\item{Step 3}\\
Finally, one has to introduce the invariant time $\tau$ and express all the Goldstone fields $u, z_n$ in terms of $t[\tau]$ and its derivatives. Moreover, these expressions have to be invariant with respect to the whole subgroup of the Virasoro group \p{Vir}.
All these can be achieved by imposing  the following constraints that generalize the constraints \p{con1}
\be\label{con2}
\omega_{-1} = d\tau, \quad \omega_n =0, \quad n \geq 0 .
\ee
\end{itemize}
As a result of  constraints \p{con2}, we obtain the following expressions for the parameters $u, z_n$
\bea\label{sol1}
u & = & Log( t' ) , \nn \\
z_1 &= & \frac{t''}{t'} = T_t, \nn \\
z_2 & = & \frac{t'''}{t'} - \frac{3}{2} \left( \frac{t''}{t'}\right)^2 = S_t, \nn \\
z_3 & = & S_t ', \nn \\
z_4 & = & S_t''-S_t^2  , \nn \\
z_5 & = & S_t^{(3)} - 2 S_t S_t' , \nn \\
z_6 & = &  S_t^{(4)} - 2 S_t S_t'' - \frac{9}{2} (S_t')^2- \frac{10}{3} S_t^3 , \nn \\
z_7 & = & S_t^{(5)} - 2 S_t S_t^{(3)} -11 S_t' S_t''-10 S_t^2 S_t', \quad \; etc .
\eea
One can compare our set of higher Schwarzians in \p{sol1} with Aharonov's invariants \cite{aharonov} and Tamanoi Schwarzians \cite{tamanoi}
\bea\label{AhTa}
\mbox{Aharonov's invariants} \; \psi_k & \mbox{     }  & \mbox{Tamanoi's Schwarzians}\; s_k \nn \\
\frac{t'[\tau]}{(t[\tau+w]-t[\tau])} = \frac{1}{w} -\sum_{k=0} \psi_{k+1} \frac{w^k}{(k+2)!} && 
\frac{t'[\tau] (t[\tau+w]-t[\tau])}{\frac{1}{2} t''[\tau](t[\tau+w]-t[\tau])+t'[\tau]^2} =
\sum_{k=0} s_k \, \frac{w^{k+1}}{(k+1)!}  \nn \\ 
\psi_1  =   \frac{t''}{ t'} =T_t && s_0 =1, \; s_1=0 \nn \\
\psi_2  =  S_t  && s_2 =  S_t \nn \\
\psi_3 = S_t' && s_3 = S_t' \nn \\
\psi_4 = S_t''+\frac{2}{3} S_t^2 & & s_4 = S_t''+4 S_t^2 \nn\\
\psi_5 = S_t'''+ 3 S_t\, S_t'  & & s_5 = S_t'''+ 13 S_t S_t'  \nn\\
\psi_6 =S_t^{(4)}+5 S_t S_t''+\frac{17}{4} ( S_t')^2 +\frac{4}{3} S_t^3 && 
s_6 = S_t^{(4)}+19 S_t''\, S_t+ 13 (S_t')^2+ 34 S_t^3 \nn \\
\psi_7 =S_t^{(5)}+\frac{22}{3} S_t S_t^{(3)}+17  S_t' S_t'' +\frac{40}{3} S_t^2 S_t' && 
s_7 = S_t^{(5)}+26  S_t^{(3)}\, S_t+ 45 S'_t S_t''+228 S_t^2 S_t' . 
\eea

Thus, we see that the difference in Schwarzians appears already at the fourth order. The same story happened in the case of Bonora-Matone Scwarzians \cite{BM}:
\bea \label{BMS}
& \mbox{ Bonora-Matone Schwarzians    } S_{2 k+1} & \nn \\
&S_{2 k+1} = \frac{1}{k} (t'[\tau])^k \partial_\tau \left( \frac{1}{t'[\tau]} \partial_\tau \left( \frac{1}{t'[\tau]} \ldots \partial_\tau (t'[\tau])^k \right)\right) & \nn \\
&S_2=S_t & \nn \\
&S_3 =S_t' & \nn \\
&S_4 = S_t''+\frac{3}{2} S_t^2 &\nn \\
&S_5 = S_t'''+8 S_t S_t' &\nn \\
& S_6 = S_t^{(4)}+\frac{31}{2} S_t S_t''+13 (S_t')^2 +\frac{45}{4} S_t^3 &\nn \\
&S_7 = S_t^{(5)}+26 S_t S_t'''+59 S_t'S_t''+144 S_t^2 S_t' & etc .
\eea

Despite the fact that we did not know the generic expressions for
the Goldstone bosons $z_n$, we  can relate $z_n$ with $\psi_n, s_n$ and/or $ S_n$. Several first such relations read:
\bea\label{zpsis}
z_2 & = & \psi_2 = s_2 =S_2, \nn \\
z_3 & = & \psi_3 = s_3 =S_3 , \nn \\
z_4 & = & \psi_4-\frac{5}{3} \psi_2^2 = s_4- 5 s_2^2 = S_4-\frac{5}{2} S_2^2, \nn \\
z_5 & = &  \psi_5- 5 \psi_2 \psi_3 =s_5-15 s_2 s_3 = S_5 - 10 S_2 S_3, \nn \\
z_6 & = & \psi_6-\frac{35}{4} \psi_3^2-7 \psi_2 \psi_4 = s_6 -\frac{35}{2} s_3^2- 21 s_2 s_4 +\frac{140}{3}s_2^2 = S_6 -\frac{35}{2} S_2 S_4- \frac{35}{2} S_3^2+\frac{35}{3} S_2^3, \nn \\
z_7 & = & \psi_7 -28 \psi_3 \psi_4 -\frac{28}{3} \psi_2 \psi_5 + \frac{70}{3} \psi_2^2 \psi_3 = 
s_7-56 s_3 s_4-28 s_2 s_5 + 350 s_2^2 s_3 = \nn \\
&& = S_7-28 S_2 S_5 -70 S_3 S_4 +175 S_3 S_2^2, \quad etc.
\eea
This means that if we insert in the parametrization of the group element $g$ \p{g} the expressions for $z_n$ in terms of $\psi_n,s_n$ and/or $S_n$, then the same constraints \p{cf1} lead to proper expressions \p{AhTa}. One should be note that  nonlinear expressions in \p{zpsis} include only $\psi_n, \psi_n$ and/or $S_n$
but not their derivatives. So, in a some sense, these are canonical transformations of variables, although non-linear.

To complete this Section, note that our constraints \p{con2} are invariant with respect to the whole Virasoro algebra \p{Vir}. If we consider the left multiplication of our group element $g$
\p{g} by the element $g_n = e^{a_n L_n}$
\be
g_n g = g'\;,
\ee
 we will obtain
\be
\delta_n t  = a_n t^{n+1}, \quad  \delta_n S_t = a_n n (n^2-1) Y_n, \quad
\delta Y_k = a_n (k+ 2 n) Y_{k+n} ,
\ee 
where
\be
Y_k = t^{k-2} (t')^2 .
\ee
It is clear that the variation of the higher Schwarzians will provide us with a new set of expressions that will  contain both $S$ and $Y^n$ and their derivatives. The usefulness of this full Virasoro symmetry is not clear for us yet.

\setcounter{equation}{0}
\section{$SL(2,\mathbb{R})$ non-invariant Schwarzians}
In addition to the Aharonov \cite{aharonov}, Tamanoi \cite{tamanoi} and Bonora-Mattone \cite{BM} versions of higher Schwarzians, 
Schippers \cite{schippers} and Bertilsson \cite{bert} proposed another definition of Schwarzian derivatives of higher order.  They have nice properties, but however they do not
possess $SL(2,\mathbb{R})$ invariance. These new higher Schwarzians are defined as follows.  \\

\noindent{\bf Schippers set}

The Schippers set of higher Schwarzians is defined as follows \cite{schippers}:

\be\label{def1}
\sigma_n:= \sigma_{n-1}' -(n-2) \frac{t''}{t'} \, \sigma_{n-1}, \quad \sigma_3 = S_t .
\ee
A few nontrivial Schwarzians have the following form:
\bea \label{SSch}
\sigma_3 & = & S_t, \nn \\
\sigma_4 & = & S'_t - 2 T_t S_t , \nn \\
\sigma_5 & = & S_t''- 2 S_t^2 - 5 T_t S_t' + 5 T_t^2 S_t, \nn \\
\sigma_6 & = & S_t^{(3)}- 9 S_t' S_t - 9 T_t S_t''+18 T_t S_t^2+\frac{45}{2} T_t^2 S_t'- 15 T_t^3 S_t, \nn \\ 
\sigma_7 & = & S_t^{(4)}-18 S_t'' S_t - 9 (S_t')^2+18 S_t^3 -14 T_t S_t^{(3)}+ 126 T_t S_t S'_t+63 T_t^2 S_t'' -126 T_t^2 S_t^2- \nn \\
& & 105 T_t^3 S'_t+\frac{105}{2} T_t^4 S_t,  \nn \\
\sigma_8 & = &  S_t^{(5)}- 32 S_t''' S_t -36 S_t' S''_t+180 S_t^2 S_t' - 20 T_t S^{(4)}+360 T_t S_t S''_t+180 T (S_t')^2-360 T_t S_t^3 + \nn \\
&& 140T_t^2 S^{(3)}-1260 T_t^2 S_t S'_t-420 T_t^3 S_t''+840 T_t^3 S_t^2+525 T_t^4 S_t' -210 T_t^5 S_t, 
\qquad etc.
\eea

\noindent{\bf Bertilsson set}\\

The Bertilsson variant of higher Schwarzians is defined as follows \cite{bert}:\\

\be\label{def2}
\cS_n:= -\frac{2}{n} ( t' )^\frac{n}{2} \frac{\partial^{n+1}}{\partial \tau^{n+1}}(t')^{-\frac{n}{2}} .
\ee

Several first members of this set read
\bea \label{BSch}
\cS_0 & = &T_t, \nn \\
\cS_1 & = & S_t , \nn \\
\cS_2 & = & S_t'- 2 T_t S_t, \nn \\
\cS_3 & = & S_t''-\frac{7}{2} S_t^2- 5 T_t S'_t+ 5 T_t^2 S_t , \nn \\
\cS_4 & = & S_t^{(3)}- 17 S_t' S_t- 9 T_t S''_t+\frac{45}{2}T_t^2 S_t'+ 34 T_t S_t^2-15 T_t^3 S_t, \nn \\
\cS_5 & = &  S_t^{(4)} - \frac{67}{2} S_t'' S_t - 22 (S_t')^2 +\frac{241}{4} S_t^3 - 14 T_t S_t^{(3)}+\frac{511}{2} T_t S_t S'_t+ 63 T_t^2 S_t''- \frac{511}{2} T_t^2 S_t^2 - \nn \\
&& 105 T_t^3 S_t' +\frac{105}{2}T_t^4 S_t, \nn \\
\cS_6 & = &  S_t^{(5)} -58 S_t^{(3)} S_t -95 S_t' S_t''+ 676 S_t^2 S_t'- 20 T_t S^{(4)}+712 T_t S_t S_t''+475 T_t (S_t')^2- 1352 T_t S_t^3+ \nn\\
&& 140 T_t^2 S^{(3)}-2730 T_t^2 S_t S'_t-420 T_t^3 S_t''+ 1820 T_t^3 S_t^2+525 T_t^4 S_t'-210 T_t^5 S_t, \qquad etc.
\eea

\subsection{Our modified set}
It is completely clear from the explicit form of the Schippers \p{SSch} and Bertilsson \p{BSch} Schwarzians that $SL(2,\mathbb{R})$ symmetry breaking is caused  by the presence  of the pre-Schwarzian derivative $T_t={\ddot t}/{\dot t}$ in these versions of Schwarzians. Within our approach, we can simulate such behaviour as follows:
\begin{itemize}
	\item{Step 1}\\
	We choose the following  parametrization of the Virasoro group element
	\be\label{hg}
	\hat{g}= e^{\im\, t L_{-1}} e^{\im\, u L_0}\;  \prod_{i=2}^{\infty} e^{\frac{\im}{(i+1)!} \hz_i L_i}
	\; e^{\frac{\im}{2}\, \hz_1 L_1} .
	\ee
	The order of the  exponents forced breaking of the invariance of  higher Goldstone fields
	$\hz_k , k\geq 2$ under transformations generated by $L_1$ and, therefore, under $SL(2,\mathbb{R})$.
		\item{Step 2}\\
	Next, one has to calculate the Cartan forms
	\be\label{cf3}
	\Omega = {\hat g}^{-1} d {\hat g} = \im\; \sum_{n=-1} \omega_n L_n .
	\ee
	\item{Step 3}\\
	Finally, one has to introduce  invariant time $\tau$ and impose constraints on the Cartan forms \p{con2}
	\be\label{con3}
	{\hat\omega}_{-1} = d\tau, \quad {\hat\omega}_n =0, \quad n \geq 0 .
	\ee
\end{itemize}
As a result, we will obtain the following expressions for the parameters $u, \hz_n$:
\bea\label{hz}
u & = & Log( t' ) , \nn \\
\hz_1 & = &T_t, \nn \\
\hz_2 & = & S_t , \nn \\
\hz_3 & = & S_t'- 2 T_t S_t, \nn \\
\hz_4 & = & S_t''- S_t^2- 5 T_t S'_t+ 5 T_t^2 S_t , \nn \\
\hz_5 & = & S_t^{(3)}- 2 S_t' S_t- 9 T_t S''_t+ 4 T_t S_t^2+\frac{45}{2}T_t^2 S_t'-15 T_t^3 S_t, \nn \\
\hz_6 & = &  S_t^{(4)} -2 S_t'' S_t - \frac{9}{2} (S_t')^2 - \frac{10}{3} S_t^3- 14 T_t S_t^{(3)}+28 T_t S_t S'_t+ 63 T_t^2 S_t''- 28 T_t^2 S_t^2 - \nn \\
&& 105 T_t^3 S_t' +\frac{105}{2}T_t^4 S_t, \nn \\
\hz_7 & = &  S_t^{(5)}- 10 S_t^2 S_t'  - 11 S_t' S_t'' -2 S_t^{(3)} S_t'- 20 T_t S^{(4)}+ 40 T_t S_t S_t''+55 T_t (S_t')^2 +20 T_t S_t^3+ \nn\\
&& 140 T_t^2 S^{(3)}-210 T_t^2 S_t S'_t-420 T_t^3 S_t''+ 140 T_t^3 S_t^2+525 T_t^4 S_t'-210 T_t^5 S_t, \qquad etc.
\eea

Thus, we see that changing  the parametrizaion of the same coset \p{hg} leads to the transformation of $SL(2,\mathbb{R})$ invariant Schwarzians \p{sol1} into non-invariant Schwarzians \p{hz}.
It should be noted here that these three sets of Schwarzians, \p{SSch}, \p{BSch} and \p{hz}, are related quite similarly to those in \p{zpsis}
\bea\label{hzsigmacs}
\hz_2 & = & \sigma_3 = \cS_1, \nn \\
\hz_3 & = &  \sigma_4 = \cS_2, \nn \\
\hz_4 & = & \sigma_5+\sigma_3^2 = \cS_3+\frac{5}{2} \cS_1^2, \nn \\
\hz_5 & = & \sigma_6 + 7 \sigma_3 \sigma_4 = \cS_4+ 15 \cS_1 \cS_2, \nn \\
\hz_6 & = & \sigma_7 + 16 \sigma_3 \sigma_5+\frac{9}{2} \sigma_4^2+\frac{32}{3}\sigma_3^3 = \cS_5+\frac{63}{2} \cS_1 \cS_3+\frac{35}{2}\cS_2^2+\frac{140}{3} \cS_1^3, \nn \\ 
\hz_7 & = & \sigma_8+25 \sigma_4 \sigma_5 +30 \sigma_3 \sigma_6+ 130 \sigma_3^2 \sigma_4 =
\cS_6+84 \cS_2 \cS_3+56 \cS_1 \cS_4+650 \cS_1^2 \cS_2, \qquad etc.
\eea
It is important that non-linear transformations from one set to another  include only Schwarzians themselves, without any derivative. Thus, one can claim that the three series of Schwarzians generate the same ring of differential operators.

\setcounter{equation}{0}
\section{Conclusion}
In this paper, a physical view was proposed on the origin of higher Schwarzians, treating them as Goldstone fields associated with the generators of the Virasoro algebra. The motivation for such an association comes from the transformation properties of higher Schwarzians under Virasoro symmetry. Using this fact, the standard nonlinear realization of the Virasoro symmetry was constructed, equipped with the constraints that did the following:
\begin{itemize}
	\item The first constraint
	$$\omega_{-1} = d \tau $$
	introduced new time $\tau$, completely inert with respect to Virasoro symmetry
	(this step is mainly the same as  in the papers \cite{AG1,AG2,AG3,KK1,KK2,KK3})
	\item Next we imposed the constraints that nullified all Cartan forms of the Virasoro group
	$$ \omega_n =0, \quad n\geq 1 $$
	These constraints express all Goldstone fields of our nonlinear realization in terms of
	the unique field $t(\tau)$ (associated with the generator $L_{-1}$) and its derivatives. The expressions for higher Goldstone fields provide a new set of higher Schwarzians.
\end{itemize}
We still have no generic expressions for our higher Schwarzians. Instead of, we presented the relations of our few first	Schwarzians  \p{sol1} with those  from the Aharonov, Tamanoi and Bonora-Matone sets \p{zpsis}.

We also explicitly demonstrated that  minor change in the coset space parametrization:
$$
 e^{\im\, t L_{-1}} e^{\im\, u L_0}\;  e^{\frac{\im}{2}\, z_1 L_1} \prod_{i=2}^{\infty} e^{\frac{\im}{(i+1)!} z_i L_i} \quad \Rightarrow \quad 
  e^{\im\, t L_{-1}} e^{\im\, u L_0}\;  \prod_{i=2}^{\infty} e^{\frac{\im}{(i+1)!} z_i L_i}  e^{\frac{\im}{2}\, z_1 L_1}
$$
together with the same constraints \p{con1} leads to $SL(2,\mathbb{R})$ non-invariant Schwarzians \p{hz}. We established the relations of a few first such Schwarzians with 
Schippers's and Bertilsson's Schwarzians \p{hzsigmacs}.

Thus, all basic variants of  higher Schwarzians have deep relations with a set of Schwarzians that follow from a nonlinear realization of Virasoro symmetry. It is clear that the construction of supersymmetric generalizations of higher Schwarzians becomes now almost straightforward along the line similar to that considered in \cite{AG2, AG3, GK, KK1,KK2,KK3}. But firstly, one has to solve the purely technical task of finding  generic expressions for our higher Schwarzians.

Finally, note that in the recent paper two possibility to retrieve the Bertilsson 
higher Schwarzians \cite{bert} from the physical systems have been discussed. However, there is no direct relation between this paper and our, because the $\ell$-conformal Galilei symmetry considered there acts, in our formulation, on the variable $\tau$ leaving all Schwarzians invariant.

\section*{Acknowledgments}
This research was partially funded by the Ministry of Science and Higher Education of Russia, Government Order for 2023-2025, Project No. FEWM-2023-0015 (TUSUR).


\begin{thebibliography}{99}
\addtolength{\itemsep}{-3pt}
\bibitem{aharonov} D.~Aharonov, {\it A nesserary and sufficient condition for univalence of a meromorphic function},\\ Duke Math. J. {\bf 36}(1969)599.
\bibitem{tamanoi} H.~Tamanoi, {\it Higher Schwarzian operators and combinatorics of the Schwarzian derivatives}, \\ Math. Ann {\bf 305} (1996) 127.
\bibitem{BM} L.~Bonora and M.~Matone,
{\it KdV equation on Riemann surfaces}, \\
Nucl.~Phys. B{\bf 327} (1989) 41; \\
M.~Matone, {\it Uniformization theory and 2-D gravity. 1. Liouville action and intersection numbers},
Int.~J.~Mod.~Phys. A{\bf 10} (1995) 289; {\tt arXiv:hep-th/9306150[hep-th]}.
\bibitem{schippers} E.~Schippers,{\it Distortion theorems for higher order Schwarzian derivatives of univalent functions},\\ Proc. of American Mathematical Society, {\bf 128} (2000) 3241.
\bibitem{bert} D.~Bertilsson, {\it Coefficient estimates for negative powers of the derivative of univalent functions}, \\
Ark. Mat. {\bf 36} (1998) 255.
\bibitem{OT} V.~Ovsienko, S.~Tabachnikov, {\it What is the Schwarzian Derivative?},\\
Notices of the AMS {\bf 56} (2009) 34.
\bibitem{nehari} Z.~Nehari, {\it The Schwarzian derivative and schlicht functions"}, 
Bull.~Amer.~Math.~Soc., {\bf 55} (1949) 545.
\bibitem{kim} S.-A~Kim, T.~Sugawa, {\it Invariant Schwarzian derivatives of higher order},\\
Complex Anal. Oper. Theory {\bf 5} (2011)659. 
\bibitem{IO} E.A.~Ivanov, V.I.~Ogievetsky, {\it The inverse Higgs phenomenon in
	nonlinear realizations},\\ Theor.~Math.~Phys. {\bf 25} (1975) 1050.
\bibitem{AG1} A.~Galajinsky, {\it Schwarzian mechanics via nonlinear realizations}, \\
Phys.Lett. B {\bf 795} (2019)277; {\tt arXiv:1905.01935[math-ph]}.
\bibitem{AG2} A.~Galajinsky, {\it Super-Schwarzians via nonlinear realizations}, \\
JHEP~{\bf 06}~(2020)~027; {\tt arXiv:2004.04489[hep-th]}.
\bibitem{AG3} A.~Galajinsky, {\it N = 3 super–Schwarzian from $OSp(3|2)$ invariants}, \\
Phys.Lett. {\bf B} 811 (2020) 135885; {\tt arXiv:2009.1306}.
\bibitem{GK} A.~Galajinsky, S.~Krivonos, {\it N = 4 super–Schwarzian derivative via nonlinear realizations},\\ Phys.Rev.  D {\bf 102} (2020)106015; {\tt arXiv:2007.04015}.
\bibitem{KK1}  N.~Kozyrev, S.~Krivonos, {\it Generalized Schwarzians},\\
Phys.Rev.  D {\bf 107} (2023)2, 026018; {\tt arXiv:2211.14021[hep-th]}.
\bibitem{KK2} N.~Kozyrev, S.~Krivonos,
{\it (Super)Schwarzian mechanics},\\
  JHEP {\bf 03} (2022)120; {\tt arXiv:2111.04643[hep-th]}.
\bibitem{KK3}  N.~Kozyrev, S.~Krivonos, {\it N=4  supersymmetric Schwarzian with $D(1,2;\alpha)$ symmetry},\\  Phys.Rev.  D {\bf 105} (2022)8, 085010; {\tt arXiv:2112.14481}.
\bibitem{AG4}  A.~Galajinsky, {\it Remarks on higher Schwarzians}, \\
Phys.~Lett.~B{\bf 843}~(2023)~138042; {\tt arXiv:2302.00317[hep-th]}. 
\end{thebibliography}
\end{document}